\documentclass[aps,apl,twocolumn,groupedaddress,showpacs,amsmath]{revtex4}
\usepackage[latin1]{inputenc}    
\usepackage{graphicx}   

\begin{document}
\title{Counting statistics and super-Poissonian noise in a quantum dot:\\
Time-resolved measurements of electron transport}
\author{S.~Gustavsson}
\email{simongus@phys.ethz.ch}
 \author{R.~Leturcq}
 \author{B.~Simovi\v{c}}
 \author {R.~Schleser}
 \author {P.~Studerus}
 \author{T.~Ihn}
 \author{K.~Ensslin}
 \affiliation {Solid State Physics Laboratory, ETH Z\"urich, CH-8093 Z\"urich,
 Switzerland}
 \author {D.C.~Driscoll}
 \author{A.C.~Gossard}
 \affiliation {Materials Departement, University of California, Santa Barbara,
 California 93106}

\date{\today}

\begin{abstract}
We present time-resolved measurements of electron transport through
a quantum dot. The measurements were performed using a nearby
quantum point contact as a charge detector. The rates for tunneling
through the two barriers connecting the dot to source and drain
contacts could be determined individually. In the high bias regime,
the method was used to probe excited states of the dot. Furthermore,
we have detected bunching of electrons, leading to super-Poissonian
noise. We have used the framework of full counting statistics (FCS)
to model the experimental data. The existence of super-Poissonian
noise suggests a long relaxation time for the involved excited
state, which could be related to the spin relaxation time.
\end{abstract}

\maketitle

\section{Introduction}
Studies of current fluctuations in conductors are of great interest
because they give information about the charge carriers in the
system and their mutual interactions, complementary to that obtained
by the measurement of the average current \cite{blanter:00}.
In recent years, the method of full counting statistics
\cite{levitov:96} (FCS) has brought renewed interest to the field.
Using  FCS, fluctuations are studied by counting the number of
electrons that pass through a conductor within a fixed period of
time. Since this gives direct access to the distribution function of
the fluctuations, not only the shot noise but also higher order
correlations can be extracted.
The method has so far mainly been used as a theoretical tool for
calculating the shot noise in various mesoscopic systems
\cite{blanter:2005}.

For electron transport through quantum dots, the noise is typically
of sub-Poissionan nature. This is due to the Coulomb blockade, which
enhances the correlation between electrons and thereby reduces the
noise \cite{davies:92}. However, when several channels with
different coupling strengths contribute to the electron transport,
interactions can lead to more complex processes and to an
enhancement of the noise \cite{sukhorukov:2001, cottet:2004,
belzig:05, onac:06}. Moreover, it has been predicted that entangled
electrons could lead to super-Poissonian noise, thus providing a
possible way of detecting entanglement in mesoscopic systems
\cite{loss:00, saraga:03}.

Experimentally, direct observations of FCS by counting electrons are
difficult to achieve. This is because a very sensitive,
non-invasive, high bandwidth charge detector is needed in order to
be able to resolve individual electrons \cite{LuW:03, fujisawa:04,
bylander:05}. Only very recently, measurements of FCS for single
level transport through a quantum dot (QD) were performed
\cite{gustavsson:05}. A quantum point contact (QPC) was used to read
out the charge state of the nearby QD \cite{field:93}. Here, we
present further time-resolved measurements of a QD system. We show
methods for tuning the QD and for extracting information about
tunneling rates \cite{schl:04} and about excited states of the QD
\cite{elz:04}.
Furthermore, we present measurements in a regime where transport is
governed by more complex processes than tunneling through a single
QD level. We observe bunching of electrons and super-Poissonian
noise. In this regime, we show that the theory of FCS
\cite{levitov:96} can be used to model the experimental data and to
extract intrinsic properties of the mesoscopic system, such as the
relaxation time between excited states.

\section{Experimental setup}
\begin{figure}[htb]
\centering
 \includegraphics[width=\columnwidth]{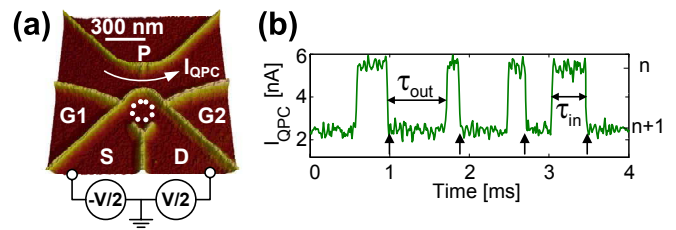}
 \caption{(Color online) (a) Quantum dot with integrated charge read-out used in the experiment.
 (b) Current through the QPC as a function of time, showing a few
 electrons tunneling into and out off the dot. The arrows mark the steps corresponding to
 an electron entering the dot.
 } \label{fig:dot}
\end{figure}

The QD used in the experiment is shown in Fig. \ref{fig:dot}(a). The
structure was fabricated using scanning probe litho\-graphy
\cite{fuhrer:04} on a $\mathrm{GaAs/Al_{0.3}Ga_{0.7}As}$
heterostructure with a two-dimensional electron gas (2DEG) 34 nm
below the surface (electron density $4.5 \times
10^{-15}~\mathrm{m}^{-2}$, mobility $25~\mathrm{m}^2 /\mathrm{Vs}$).
The sample consists of a QD [dotted circle in Fig. \ref{fig:dot}(a)]
and a nearby QPC. We estimate from the geometry and the
characteristic energy scales that the dot contains about $30$
electrons. The gates $G1$ and $G2$ were used to tune the tunnel
barriers connecting the dot to source and drain leads, while the $P$
gate was used to tune the conductance of the QPC to a regime where
the sensitivity to changes in the dot charge is maximal. For our
setup, the best sensitivity was reached when the QPC conductance
($G_{QPC}$) was tuned below the first conductance plateau, with
$G_{QPC} \sim 0.25 \, e^2/h$. The conductance was measured by
applying a voltage over the QPC ($V_{\mathrm{bias}} =
500~\mathrm{\mu V}$) and monitoring the current. Since changing the
voltages on gates $G1$ and $G2$ also affects the QPC sensitivity, a
compensation voltage had to be applied to the $P$-gate in order to
keep the QPC in the region of maximum sensitivity whenever the other
gates were changed. All measurements where performed in a dilution
refrigerator with a base temperature of 60 mK.

When an electron tunnels onto the dot, the conductance through the
QPC is reduced due to the electrostatic coupling between the dot and
the QPC. A typical time trace of the QPC current is plotted in Fig.
\ref{fig:dot}(b), showing switching between two levels. The low
levels correspond to the configuration where the dot contains one
extra electron, while $\tau_{\mathrm{in}}$ and $\tau_{\mathrm{out}}$
specify the time it takes for an electron to tunnel into and out of
the dot, respectively. The length of each time trace presented here
is 0.5 s.

The bandwidth of the QPC circuit is $\Delta f = 30~\mathrm{kHz}$,
which limits the current we can measure by counting electrons to $I
\leq e \Delta f \sim 5~\mathrm{fA}$. The bandwidth is similar to
what was achieved in measurements on a split-gate defined dot
\cite{vand:04}. In our setup, the bandwidth is not limited by low
signal-to-noise ratio (S/N), but by the low-pass filter formed by
the cable capacitance and the feedback resistor of the IV-converter.
From the trace shown in Fig. \ref{fig:dot}(b), we extract
$\mathrm{S/N} \sim 15$. Assuming a flat noise spectrum, we estimate
that S/N would allow us to increase the bandwidth by a factor of ten
and still get a detectable signal. One possible reason for the high
sensitivity of our detector compared to split-gate defined
structures is that there are no metallic gates on the surface that
shield the electrostatic coupling between the QD and the QPC.
However, the sensitivity is also strongly dependent on the exact
shape of the confinement potential within the QPC and on how
susceptible this potential is to changes in the electrostatical
environment. This may vary a lot from sample to sample. For the
structure used in this measurement, the QPC showed broad resonances
in addition to the standard plateau features. By operating the QPC
at the flank of a resonance in the step below the first plateau, we
were able to find a regime with good sensitivity. For typical
current levels in the QPC (nA), electrons pass the QPC at a rate
which is many orders of magnitude higher than the rate for electrons
passing the quantum dot (aA). We conclude that back action of the
QPC via its shot noise can be neglected for the analysis of the
counting statistics \cite{gurvitz:97}.

\section{Thermal noise with one lead connected to the dot}
In the following, we are interested in the number of electrons
visiting the dot during a given time interval. We call each visit
one \emph{event} and use the symbol $r_E$ to denote the number of
events occurring per second. In the low-bias Coulomb blockade
regime, the dot can only hold one excess electron. Before a new one
can enter, another one has to go out. In this case, we can count the
events by detecting the electrons as they enter the dot [marked by
vertical arrows in Fig. \ref{fig:dot}(b)]. Note that by counting
events, we do not distinguish between electrons passing through the
dot and electrons hopping back and forth between the dot and a
single lead.

First, we concentrate on the regime where only one lead is connected
to the dot and the electron motion is entirely governed by thermal
fluctuations and occupation probabilities.  For the data shown in
Fig. \ref{fig:dot}(b), the gates are tuned such that the tunnel
barrier between the dot and the drain lead is completely closed,
while the source lead is weakly coupled to the dot. With only one
lead open and with a temperature and level broadening much lower
than the charging energy and single level spacing of the QD, only
one QD state is available for tunneling. The probability for an
electron to tunnel into or out of the dot during a time interval dt
is governed by the relation
\begin{equation}\label{eq:expDecay}
p_{\mathrm{in/out}}(t) \mathrm{dt} = \Gamma_{\mathrm{in/out}}
\mathrm{e}^{-\Gamma_{\mathrm{in/out}} t} \mathrm{dt},
\end{equation}
where $\Gamma_{\mathrm{in}}$ and $\Gamma_{\mathrm{out}}$ are the
effective rates for tunneling into and out of the dot. Using similar
methods as in Ref. \cite{schl:04,gustavsson:05}, we have checked
that Eq. \ref{eq:expDecay} is fulfilled when we are in the
single-level regime. In the following, we consider the case of a
non-degenerate level, as discussed in previous work \cite{schl:04}.
To unambiguously determine the spin configuration of the involved
states, measurements as a function of magnetic field would be
required. Given the availability of our experimental data we focus
our analysis on the extraction of the tunneling rates for a single
non-degenerate state. We also assume the tunnel coupling to be
independent of energy within the small interval of interest. At the
end of this section we discuss how the analysis would change if spin
degeneracy is relevant.

The relations between the effective rates and the dot-lead tunnel
coupling $\Gamma$ are given by
\begin{equation}\label{eq:oneLeadRate}
 \Gamma_{\mathrm{in}} = \Gamma \, f(\Delta E/k_B T) ,~ \Gamma_{\mathrm{out}}
 = \Gamma \, (1-f(\Delta E/k_B T)),
\end{equation}
where $f(x)$ is the Fermi distribution function, $T$ is the
temperature and $\Delta E$ is the energy difference between the
Fermi level of the lead and the electrochemical potential of the
dot.

The tunneling rates can be determined directly from the measured
time traces. Using Eq. \ref{eq:expDecay}, we find
$\Gamma_{\mathrm{in}} = 1/\langle \tau_{\mathrm{in}} \rangle,
~\Gamma_{\mathrm{out}} = 1/\langle \tau_{\mathrm{out}} \rangle$,
with a relative accuracy of $\sqrt{2 \ln 2 /N}$ (see Appendix
\ref{app:rates}). Here, $N$ is the total number of switches
occurring during one trace. The relative accuracy is calculated
assuming that Eq. \ref{eq:expDecay} is valid. For the trace in Fig.
\ref{fig:dot}(b), we get $\Gamma_{\mathrm{in}} = 8.0~\mathrm{kHz}$,
$\Gamma_{\mathrm{out}} = 2.3~\mathrm{kHz}$ and $\Gamma = 1/\langle
\tau_{\mathrm{in}}\rangle + 1/\langle \tau_{\mathrm{out}}\rangle =
10.3~\mathrm{kHz}$, with a relative accuracy of $1.7\%$. It has been
shown that the finite bandwidth of the detector leads to a
systematic under-estimate of the actual rates. For the rates given
here we have compensated for such errors using the methods presented
in Ref. \cite{naaman:2006}, with a detection rate of
$\Gamma_{\mathrm{det}}=100~\mathrm{kHz}$.

\begin{figure}[htb]
\centering
 \includegraphics[width=\columnwidth]{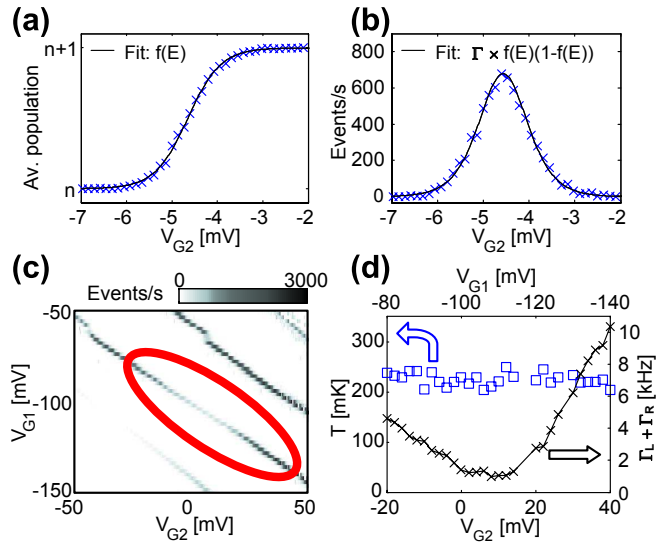}
 \caption{(Color online)
 (a) Average dot population versus voltage on gate $G2$. The data was fit to a Fermi distribution function
 with $T = 230~\mathrm{mK}$.
 (b) Counts of events per second for the same data as in (a). The
 data was fit to Eq. (\ref{eq:eventPerSec}), giving $\Gamma = 2.63~\mathrm{kHz}$ and $T = 230~\mathrm{mK}$.
 (c) Events per second versus $V_{G1}$ and $V_{G2}$. For low values
 of $V_{G1}$ and $V_{G2}$, both the source lead and the drain lead are
 pinched off. For
 high voltages, the barriers open up so much that the tunneling occurs on a timescale
 faster than the measurement bandwidth.
 (d) Temperature (squares) and tunnel coupling (crosses), extracted from data shown within the ellipse in
 (c). As $V_{G2}$ is increased, $V_{G1}$ is decreased, in order
 to keep the dot at a constant potential. For low $V_{G2}$,
 tunneling occurs between the source lead and the dot, for high
 $V_{G2}$, the electrons tunnel between the drain and the dot. For
 intermediate gate values, both leads contribute to the tunneling.
 The electron temperature was found to be the same for both leads,
 within the accuracy of the data analysis.
 } \label{fig:gammas}
\end{figure}

Since the dot can only hold one extra electron, we can determine the
Fermi function from the average population of excess electrons on
the dot
\begin{equation}\label{eq:fermiPop}
 f(\Delta E/k_B T) = \langle n_{\mathrm{excess}}\rangle = \langle \tau_{\mathrm{out}}\rangle/(\langle
 \tau_{\mathrm{in}}\rangle + \langle \tau_{\mathrm{out}}\rangle).
\end{equation}
The Fermi function can also be found by counting the average
number of events occuring per second, $r_E$. Assuming sequential
tunneling and using Eq. (\ref{eq:oneLeadRate}), we find for the
case with one lead open
\begin{equation}\label{eq:eventPerSec}
    r_E = 1/(\langle \tau_{\mathrm{in}}\rangle + \langle \tau_{\mathrm{out}}\rangle)
    = \Gamma \, f (1-f).
\end{equation}
In Fig. \ref{fig:gammas}(a) and (b) we plot the average population
and the number of events per second as the gate $G2$ was used to
change the electrochemical potential of the dot. The accuracy
obtained when determining the Fermi function is $\Delta f =
f(1-f)\sqrt{2/N}$ (see Appendix \ref{app:fermi}), giving error bars
smaller than the markers used in the figures. The data fits well to
the expected relations. By first determining the lever arm between
gate $G2$ and the dot from standard Coulomb diamond measurements
\cite{kouw:97}, it was possible to extract the electronic
temperature ($T=230~\mathrm{mK}$) from the width of the Fermi
function. The same temperature was found by checking the width of
standard Coulomb blockade current peaks \cite{kouw:97}, measured
when the dot was in a more open regime.

As mentioned earlier, the results shown here are valid only for a
non-degenerate level. Taking spin degeneracy into account will
modify the value extracted for the tunneling rate $\Gamma$ (see
Appendix \ref{app:rates}), but it will not change the width of the
Fermi distribution. The error analysis is performed within the
assumption of a non-degenerate level. If spin-degeneracy is taken
into account, then some tunneling rates are changed by a factor of
two, while the expression for the relative accuracy remains the same
as derived before. Further experiments are required to clearly
differentiate between spin-degenerate levels and single levels at
the Fermi energy. However, the general way how our analysis proceeds
is not affected by this.

\section{Thermal noise with two leads connected to the dot}
In order to perform time-resolved measurements of electron transport
through the dot, the tunnel barriers have to be symmetrized so that
both give similar tunneling rates. The rates must be kept lower than
the bandwidth of the setup, but still high enough to give good
statistics. Figure \ref{fig:gammas}(c) shows the number of events
per second as a function of the two gates $V_{G1}$ and $V_{G2}$. In
the upper left corner of the figure, $V_{G1}$ is high and $V_{G2}$
is low, corresponding to the case where the source lead is open and
the drain lead is closed. In the bottom right corner, the opposite
is true. For the region in between, marked by the ellipse in Fig.
\ref{fig:gammas}(c), the data indicates that both leads are weakly
coupled to the dot.

The measurement method does not enable us to distinguish whether an
electron that tunnels into the dot arrives from the left or from the
right lead. Therefore, when both leads are connected to the dot, the
rates in Eq. (\ref{eq:oneLeadRate}) must be adjusted to contain one
part for the left lead and one part for the right lead,
\begin{eqnarray}\label{eq:bothLeadGamma}
 \nonumber &\Gamma_{\mathrm{in}} = \Gamma^{\mathrm{in}}_L + \Gamma^{\mathrm{in}}_R =
 \Gamma_L f_L + \Gamma_R f_R, \\
 &\Gamma_{\mathrm{out}} = \Gamma^{\mathrm{out}}_L + \Gamma^{\mathrm{out}}_R =
 \Gamma_L (1-f_L) + \Gamma_R (1-f_R).
\end{eqnarray}
Here, $f_L$ and $f_R$ are the Fermi distribution functions of the
left and the right lead, respectively. Using Eq.
(\ref{eq:bothLeadGamma}), we calculate the rate of events for the
case when both leads are kept open,
\begin{equation}\label{eq:eventPerSec2}
 r_E = \frac{[\Gamma_L f_L + \Gamma_R f_R] [\Gamma_L (1-f_L) + \Gamma_R (1-f_R)]}
 {\Gamma_L + \Gamma_R}.
\end{equation}
With no bias applied to the dot, the two distributions functions
$f_L$ and $f_R$ are identical except for a possible difference in
electronic temperature in the two leads. However, assuming $T_L =
T_R = T$, we have $f_L = f_R = f$, and Eq. (\ref{eq:eventPerSec2})
simplifies to $r_E = (\Gamma_L + \Gamma_R) \, f(1-f)$. Fitting this
expression to curves similar to that shown in Fig. \ref{fig:gammas}(b),
we extract the temperature and combined tunneling rate $\Gamma_L +
\Gamma_R$ from the data within the ellipse of Fig. \ref{fig:gammas}(c).
The result is presented in Fig. \ref{fig:gammas}(d). The rates and the
temperature shown in the graph are due to the combined tunneling
through both leads. Still, for low $V_{G2}$ (high $V_{G1}$), the
drain lead is pinched off and tunneling occurs mainly between the
source lead and the dot. For high $V_{G2}$ (low $V_{G1}$), the
source is pinched off and the tunneling is dominated by electrons
going between the drain and the dot. The fact that the electronic
temperatures extracted from both regimes turn out to be the same
within the accuracy of the analysis (T=230 mK) justifies the
assumption that $T_L = T_R$.

\section{Shot noise at finite bias}
\begin{figure}[tbh]
\centering
 \includegraphics[width=\columnwidth]{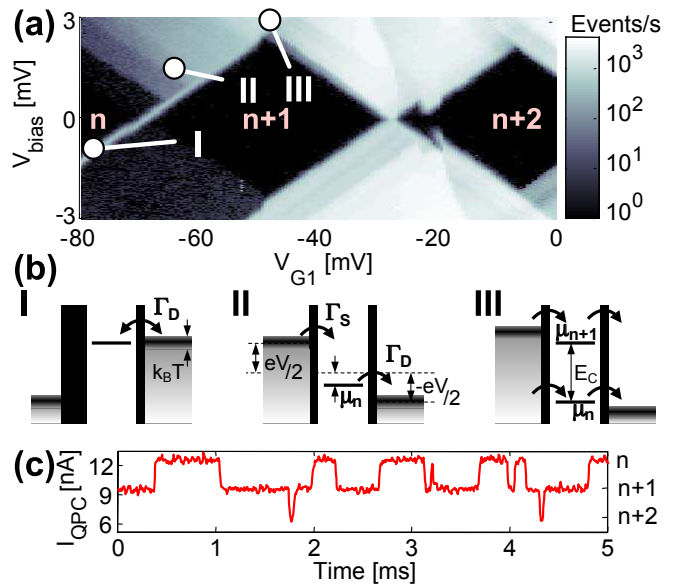}
 \caption{(Color online) (a) Coulomb diamonds, measured by counting events per second. For
 low values of $V_{G1}$, the source lead is pinched off and
 tunneling can only occur between the dot and the
 drain lead. As $V_{G1}$ increases, the source lead opens up and a current
 can flow through the dot.
 (b) Diagrams depicting the energy levels of the dot at points I, II and III. In case III,
 the bias is higher than the charging energy of the dot, meaning that
 the dot can contain 0, 1 or 2 excess electrons. (c) Time trace
 taken at point III. The three possible dot populations ($n$, $n+1$
 or $n+2$ electrons) are clearly resolvable.
 } \label{fig:cd}
\end{figure}

Now we apply a finite voltage bias between source and drain leads
and measure electron transport through the dot. Figure
\ref{fig:cd}(a) shows the Coulomb blockade diamonds measured by
counting events. In this measurement, the gate $G1$ was used as a
plunger gate to control the dot electrochemical potential. However,
the gate also strongly affects the source tunnel barrier. For low
$G1$ voltages, the source lead is closed, giving strong charge
fluctuations only when the drain lead is in resonance with the dot
[see case I in Fig. \ref{fig:cd}(a,b)].

At higher gate voltages, the source lead opens up and a current
can flow through the dot. In point II of Fig. \ref{fig:cd}(a), the
dot electrochemical potential $\mu_n$ lies within the bias window
but far away from the thermal broadening of the Fermi distribution
in the leads. The condition can be expressed as
\begin{equation}\label{eq:condCurrent}
    |\!\pm \!eV/2-\mu_n| \gg k_B T,
\end{equation}
where the "+" case refers to the source contact and the "-" case
refers to the drain. Whenever Eq. (\ref{eq:condCurrent}) is
fulfilled, electrons can only enter the dot from the source lead and
only leave through the drain. In this regime, we measure the current
through the dot by counting events. This opens the possibility to
use the QD as a very precise current meter for measuring sub-fA
currents \cite{bylander:05}. Since the electrons are detected one by
one, the noise and higher order correlations of the current can also
be experimentally investigated \cite{gustavsson:05}. In this regime
we measure the shot noise of the system, which arises because of the
discreteness of the charge carriers \cite{blanter:00}. This is in
contrast to the results shown in the previous sections, where the
fluctuations were due to thermal effects.

When the bias exceeds the dot charging energy, $E_C \sim
2.1~\mathrm{meV}$, and the electrochemical potentials of the $(n)$
and the $(n+1)$ states are within the bias window [see case III of
Fig \ref{fig:cd}(a,b)], transport processes are allowed where the
dot may contain 0, 1 or 2 excess electrons. A time trace measured at
point III of Fig. \ref{fig:cd}(a) is shown in Fig. \ref{fig:cd}(c).
The high sensitivity of the QPC charge detector allows us to measure
switching between three different levels, corresponding to $(n)$,
$(n+1)$ and $(n+2)$ electrons on the dot. This distinction is not
possible in a standard current measurement.


With the condition given by Eq. (\ref{eq:condCurrent}) fulfilled, we
know that for positive bias voltage, electrons always enter the dot
through the source contact and leave the dot through the drain
contact. In this case, we have
\begin{equation}\label{eq:gammaInSource}
 \Gamma_{S} = \Gamma_{\mathrm{in}} = 1/\langle \tau_{\mathrm{in}} \rangle, ~~~
 \Gamma_{D} = \Gamma_{\mathrm{out}} = 1/\langle \tau_{\mathrm{out}} \rangle.
\end{equation}

Equation (\ref{eq:gammaInSource}) can then be used to determine
the tunneling rates of an individual state, but only if there are
no excited states available within the bias window. If there are
excited states available, Eq. (\ref{eq:gammaInSource}) will still
be valid, however, the calculated $\Gamma_{S}$ and $\Gamma_{D}$
will not be the tunneling rates of a single state but rather the
sum of rates from all states contributing to the tunneling
process. A further complication with excited states is that there
may be equilibrium charge fluctuations between the lead and the
excited state, thereby removing the unidirectionality of the
electron motion. However, if the relaxation rate of the excited
state into the ground state is orders of magnitude faster than the
tunneling out rate, the electron in the excited state will have
time to relax to the ground state before equilibrium fluctuations
can take place.

\begin{figure}[htb]
\centering
 \includegraphics[width=\columnwidth]{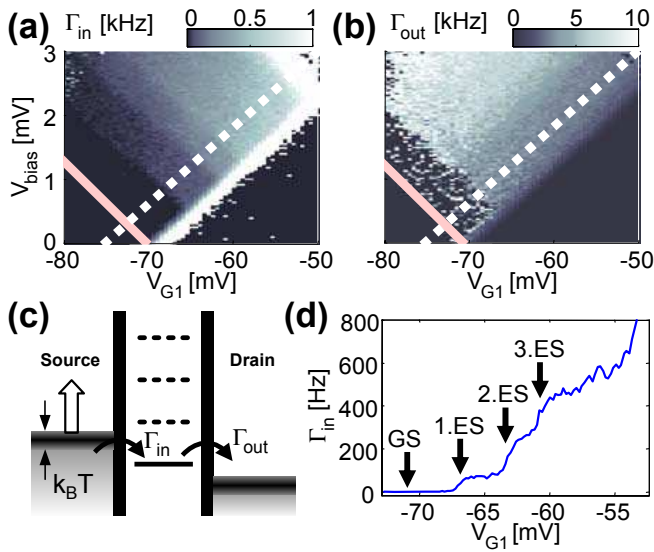}
 \caption{(Color online) (a) and (b): Blow-up of the upper left region of Fig. \ref{fig:cd}(a), showing the
 rates for electrons tunneling in (a) and out (b) of the dot,
 respectively. The solid lines mark the positions where the
 source lead is lining up with the electrochemical potential of the dot's ground
 state. The dashed lines mark the lower edge of the region where condition of Eq. (\ref{eq:condCurrent}) in the text is fulfilled. The color scales are different for the two figures,
 the rate for tunneling out is around 10 times faster than tunneling
 in.
 (c) Diagram depicting the energy levels along the dashed lines in
 (a) and (b). As the source lead is raised [corresponds to going upward
 along the dashed lines in (a) and (b)], excited states become
 available for tunneling.
 (d) Tunneling rate for electrons entering the dot, measured along
 the dashed line in (a). With increased gate voltage, excited
 states become available for transport, giving higher tunneling
 rates.
 }\label{fig:gammaInOut}
\end{figure}

The separate rates $\Gamma_{\mathrm{in}}$ and
$\Gamma_{\mathrm{out}}$ for a close-up of the upper-left region of
Fig. \ref{fig:cd}(a) are plotted in Fig. \ref{fig:gammaInOut}(a) and
(b). It is important to note that the requirement of Eq.
(\ref{eq:condCurrent}) is met only for the region along and above
the dashed lines in the figures. At the lower left end of the dashed
lines, the energy levels of the dot are aligned as shown in Fig.
\ref{fig:gammaInOut}(c). Going diagonally upward along the lines
corresponds to raising the Fermi level of the source lead, while
keeping the energy difference between the dot and the drain lead
fixed.

Starting at low bias and low voltage on the gate $V_{G1}$, the dot
is in the Coulomb blockade regime, and no tunneling is possible.
Following the dashed line upwards, the dot ground state becomes
available for tunneling at $V_{\mathrm{bias}}=0.3~\mathrm{mV}$.
The transitions is marked by the solid lines in Fig.
\ref{fig:gammaInOut}(a,b). At these low gate voltages, the source
tunnel barrier is almost completely pinched off, meaning that the
rate for electrons entering the dot is still low [Fig.
\ref{fig:gammaInOut}(a)]. Even so, some electrons do enter the
dot, as can be seen from the few points of measurements of rates
for electrons tunneling out of the dot within the corresponding
region of Fig. \ref{fig:gammaInOut}(b).

We now concentrate on the tunneling-in rate in Fig.
\ref{fig:gammaInOut}(a). As the source level is further raised,
excited states become available for transport. The first excited
state (at $V_{\mathrm{bias}}=0.85~\mathrm{mV}$ along the dashed
line) is more strongly coupled to the lead than the ground state,
giving a tunneling rate of $\sim\! 70~\mathrm{Hz}$ for electrons
entering the dot. The large difference in the tunneling-in rate
between the ground and the excited state can be understood if the
wavefunctions of the ground and excited state have different
spatial distributions. If the overlap with the lead wavefunction
is larger for the excited state, the tunneling rate will also be
larger. Similar differences in tunneling rates have been found
between the singlet and triplet states in a two-electron dot
\cite{hans:05, ciorga:00}.

By further raising the source level, tunneling can also occur
through a second excited state. The measured tunneling-in rate
will now be the sum of the rates from both excited states; by
subtracting the contribution from the first state, the
tunneling-in rate for the second state can be determined. Using
this method, we can resolve three excited states, with excitations
energies $\varepsilon_1 = 0.55~\mathrm{meV}$, $\varepsilon_2 =
1.0~\mathrm{meV} $, $\varepsilon_3 = 1.3~\mathrm{meV}$ and with
tunneling rates $\Gamma_1 = 70~\mathrm{Hz}$, $\Gamma_2 =
190~\mathrm{Hz}$, $\Gamma_3 = 190~\mathrm{Hz}$. The excited states
are clearly seen in Fig. \ref{fig:gammaInOut}(d), which is a cut
along the dashed diagonal line in Fig. \ref{fig:gammaInOut}(a).


Focusing now on the rates for electrons tunneling out of the dot
[Fig. \ref{fig:gammaInOut}(b)], there is a noisy region where the
ground state but no excited states are within the bias window
($0.3<V_{\mathrm{bias}}<0.85~\mathrm{mV}$ along the dashed line).
In this regime, few electrons will enter the dot, meaning that the
statistics needed for measuring the rate of electrons leaving the
dot is not sufficient. However, for bias voltages higher than the
first excited state, the tunneling-out rate remains constant along
the dashed line. This is in contrast to the steps seen in the
tunneling-in rates, indicating that the rate for tunneling out of
the QD does not depend on the state used for tunneling into the
QD. Since the individual excited states are expected to have
different rates also for tunneling out of the dot, the data is
consistent with the interpretation that an electron entering the
dot into an excited state will always have time to relax to the
ground state before it tunnels out. The rate for tunneling out is
$\sim\! 6~\mathrm{kHz}$, giving an upper bound for the relaxation
time of $\sim \! 170~\mathrm{\mu s}$.

The main relaxation mechanism in quantum dots is thought to be
electron-phonon scattering \cite{ino:92}. Measurements on
few-electron vertical quantum dots have shown relaxation times of
$10~\mathrm{ns}$ \cite{fuji:02}. Recent numerical investigations
have shown that the electron-electron interaction in multi-electron
dots can lead to reduced relaxation rates \cite{bert:05}. Still, the
relaxation rate is expected to be considerably faster than the upper
limit we give here.

\section{Bunching of electrons}
\begin{figure}[htb]
\centering
 \includegraphics[width=\columnwidth]{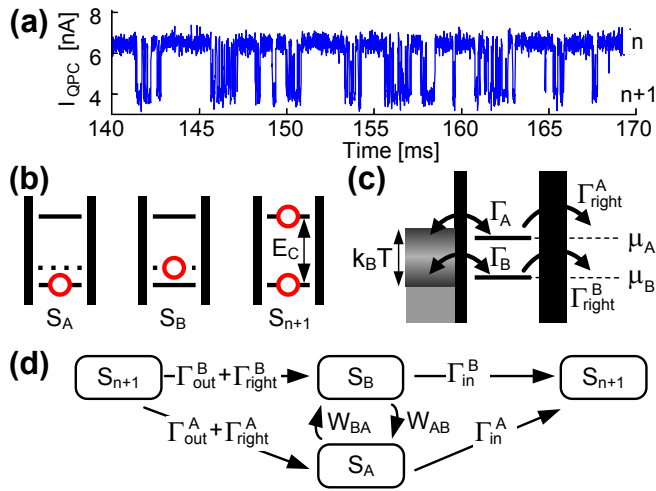}
 \caption{(Color online) (a) Time trace of the QPC current showing bunching of
 electrons. (b) Dot states included in the model used to describe the bunching
 of electrons. The red circles correspond to electron occupation.
 State $S_A$ is the $n$-electron ground state, state $S_B$ is an excited
 $n$-electron state and state $S_{n+1}$ is the ground state when
 the dot contains $(n+1)$ electrons.
 (c) Energy diagram for the model. The two dot transitions are both
 within the thermal broadening of the lead.  Electrons enter the dot from the left
 lead and may leave through either the left or the right lead.
 (d) Possible transitions between the different
 states of the model.
 The rates $\Gamma^{\mathrm{A}}_{\mathrm{in}}$, $\Gamma^{\mathrm{B}}_{\mathrm{in}}$
 refer to
 electrons entering the dot, thus taking the dot from state
 $S_{A/B}$ to state $S_{n+1}$. The rates
 $\Gamma^{\mathrm{A}}_{\mathrm{out}}$, $\Gamma^{\mathrm{B}}_{\mathrm{out}}$ describe electrons leaving
 the dot, giving transitions from state $S_{n+1}$ to $S_{A/B}$. $W_{AB}$ and
 $W_{BA}$ are the direct transition rates between states $S_A$ and $S_B$.
 Finally, the rates $\Gamma^{\mathrm{A}}_{\mathrm{right}}$, $\Gamma^{\mathrm{B}}_{\mathrm{right}}$ refer to
 electrons leaving the dot through the right lead.
 }\label{fig:spinTrace}
\end{figure}

So far, we have analyzed data where the tunneling events can be well
explained by a rate equation approach with one rate for electrons
tunneling into and another rate for electrons leaving the dot. For
the trace shown in Fig. \ref{fig:spinTrace}(a), the behavior is
distinctly different. The electrons come in bunches; there are
intervals where tunneling occurs on a fast timescale
($>\!10~\mathrm{kHz}$), in-between these intervals there are long
periods of time ($>\!1~\mathrm{ms}$) without any tunneling. The data
was taken with a bias applied so that the Fermi level of the source
lead is lining up with the electrochemical potential of the dot,
while the drain lead is far below the electrochemical potential of
the dot, thus prohibiting electrons from entering the dot from the
drain lead. The voltage on gate $V_{G1}$ was set to
$34~\mathrm{mV}$, which is outside the range of the Coulomb diamonds
presented in Fig. \ref{fig:cd}(a). Since the QPC current is at the
high level during the intervals without tunneling, the dot contains
one electron less when the fast tunneling is blocked.

In order to explain the two different timescales, we assume the
validity of a model where there are two almost energy-degenerate dot
states within the thermal broadening of the distribution in the
source lead. Because of Coulomb blockade, the dot may hold one or
zero excess electrons. The model includes three possible dot states,
shown in Fig. \ref{fig:spinTrace}(b). State $S_A$ is the
$n$-electron ground state, state $S_B$ is an excited $n$-electron
state and state $S_{n+1}$ is the ground state when the dot contains
$(n+1)$ electrons. Transitions between the $S_A$/$S_B$ states and
the $S_{n+1}$ state occur whenever an electron tunnels into or out
of the dot.

The tunnel coupling between the dot and the lead is given by the
overlap of the dot and lead electronic wavefunctions. Since the
wavefunctions corresponding to the two states $S_A$ and $S_B$ may
have different spatial distributions, the coupling strength
$\Gamma_A$ of the transition $S_A \Leftrightarrow S_{n+1}$ can vary
from the coupling $\Gamma_B$ of the $S_B \Leftrightarrow S_{n+1}$
transition. The energy levels of the dot and the leads for the
configuration where we measure bunching of electrons are shown in
Fig. \ref{fig:spinTrace}(c), while the possible transitions of the
model are depicted in Fig. \ref{fig:spinTrace}(d).

Starting with one excess electron on the dot [state $S_{n+1}$ in
Fig. \ref{fig:spinTrace}(d)], at some point an electron will tunnel
out, leaving the dot in either state $S_A$ or state $S_B$. Assuming
$\Gamma_B \gg \Gamma_A$, it is most likely that the dot will end up
in the excited state $S_B$. If the tunneling rate $\Gamma_B$ is
faster than the relaxation process $S_B \Rightarrow S_A$, an
electron from the lead will have time to tunnel onto the dot again
and take the dot back to the initial $S_{n+1}$ state. The whole
process can then be repeated, leading to the fast tunneling in Fig.
\ref{fig:spinTrace}(a).

However, at some point the dot will end up in state $S_A$, either
through an electron leaving the dot via the $\Gamma_A$ transition,
or through relaxation of the $S_B$ state. To get out of state $S_A$,
there must be either a direct transition back to state $S_B$, or an
electron tunneling into the dot through the $S_A \Rightarrow
S_{n+1}$ transition. With $\Gamma_B \gg \Gamma_A$ and assuming
$\Gamma_B \gg ~W_{BA}$, both processes are slow compared to the
tunneling between the lead and state $S_B$.
This mechanism will block the fast tunneling and produce the
intervals without switching events seen in Fig.
\ref{fig:spinTrace}(a). Similar arguments can be used to show that
the blocking mechanism will be possible also if $\Gamma_B \ll
\Gamma_A$.

From the above reasoning, we see that the fast timescale is set by
the fast tunneling state, while the slow timescale is determined
either by the relaxation process $S_B \Rightarrow S_A$ or by the
slow tunneling rate, depending on which process is the fastest.
Either way, it is crucial that the relaxation rate is slower than
the fast tunneling rate (in our case $W_{AB} \ll \Gamma_B \sim
20~\mathrm{kHz}$). We speculate that the slow relaxation rate may be
due to different spin configurations of the two states. For a
few-electron QD, spin relaxation times of $T_1>1~\mathrm{ms}$ have
been reported \cite{hans:05, elzNat:04}.

To make quantitative comparisons between the model and the data, we
use the framework of full counting statistics (FCS) to investigate
how the dot charge fluctuations change as the source lead is swept
over a Coulomb resonance. Theoretical investigations of multi-level
quantum dots have lead to predictions of electron bunching and
super-Poissonian noise \cite{belzig:05}. Following the lines of
Refs. \cite{bagrets:03, belzig:05}, we first write the master
equation for the system,
\begin{equation}\label{eq:master}
 \frac{d}{dt}
 \left( \begin{array}{l}
 p_A \\
 p_B \\
 p_{n+1}
 \end{array}\right) = M
 \left( \begin{array}{l}
 p_A \\
 p_B \\
 p_{n+1}
 \end{array}\right),
\end{equation}
with $M=$
\begin{equation}\label{eq:defM}
\left(
 \begin{array}{ccc}
 -\Gamma_{\text{in}}^A - W_{BA} & W_{AB} & (\Gamma _{\text{out}}^A+\Gamma
   _{\text{right}}^A) * e^{i\chi} \\
 W_{BA} & -\Gamma _{\text{in}}^B-W_{AB} & (\Gamma _{\text{out}}^B+\Gamma
   _{\text{right}}^B) * e^{i\chi} \\
 \Gamma _{\text{in}}^A & \Gamma _{\text{in}}^B & - \Gamma_{\text{out}}
\end{array}
\right).
\end{equation}
Here $\Gamma_{\text{out}} = (\Gamma
_{\text{out}}^A+\Gamma_{\text{out}}^B+\Gamma
_{\text{right}}^A+\Gamma_{\text{right}}^B)$ and $p_A$, $p_B$ and
$p_{n+1}$ are occupation probabilities for states $S_A$ and $S_B$
and $S_{n+1}$, respectively.
The effective tunneling rates are determined by multiplying the
tunnel coupling constants for each state with the Fermi distribution
of the electrons in the lead,
\begin{equation}\label{eq:effGamma}
 \Gamma _{\text{in/out}}^{A/B} = f[\mp (eV - \mu_{A/B})] \,
 \Gamma_{A/B}.
\end{equation}
The tunneling rates $\Gamma _{\text{right}}^A$ and $\Gamma
_{\text{right}}^B$ are included to account for the possibility for
electrons to leave through the right barrier. The Fermi level of the
right lead is far below the electrochemical potential of the dot, so
that the states in the right lead can be assumed to be unoccupied.
Finally, $W_{AB}$ and $W_{BA}$ are the direct transition rates
between states $S_A$ and $S_B$. These rates obey detailed balance,
\begin{equation}\label{eq:detBal}
W_{AB}/W_{BA} = \exp\left[(\mu_A - \mu_B)/k_B T\right].
\end{equation}
The phenomenological relaxation rate between the two states is given
as $1/T_1 = W_{AB} + W_{BA}$.

In Eq. (\ref{eq:defM}), we introduce charge counting by multiplying
all entries of $M$ involving an electron leaving the dot with the
counting factor $\exp(i\chi)$ \cite{bagrets:03}. We do not
distinguish whether the electron leaves the dot through the left or
the right lead. In this way we obtain the counting statistics
$p_{t_0}(N)$, which is the probability for counting $N$ events
within the time span $t_0$. The distribution describes fluctuations
of charge on the dot, which is exactly what is measured by the QPC
detector in the experiment. We stress that this distribution is
equal to the distribution of current fluctuations only if it can be
safely assumed that the electron motion is unidirectional. This is
the case if the condition in Eq. (\ref{eq:condCurrent}) is
fulfilled, i.e. if the tunneling due to thermal fluctuations is
suppressed. Here, we are in a regime where there is a mixture of
tunneling due to the applied bias and tunneling due to equilibrium
fluctuations. But since the model defined in Eq. (\ref{eq:defM}) is
valid regardless of the direction of the electron motion, it can
still be used for analyzing the experimental data.

Using the method of Ref. \cite{bagrets:03}, we calculate the lowest
eigenvalue $\lambda_0(\chi)$ of $M$ and use it to obtain the
cumulant generating function (CGF) for $p_{t_0}(N)$,
\begin{equation}\label{eq:sOfChi}
  S(\chi) = -\lambda_0 (\chi) t_0.
\end{equation}
The CGF can then be used to obtain the cumulants of any order using
the relation $C_n = -(-i \partial_{\chi})^n S(\chi)|_{\chi=0}$. In
order to compare the theory with the experiment we extract the first
three cumulants of $p_{t_0}(N)$ from the experimental data. Since we
want to compare the data with the predictions given by the CGF of
the model, we choose to calculate the cumulants instead of the
central moments, as it was done in a previous work
\cite{gustavsson:05}. The first cumulant ($C_1$) is identical to
$\langle N \rangle$, the mean of the distribution, while the second
and third cumulants ($C_2$, $C_3$) coincide with the second and
third central moments [$\langle N^2 \rangle - \langle N \rangle^2$
and $(N-\langle N \rangle )^3$], giving the variance and the
asymmetry of the distribution.

The cumulants were found by taking a trace of length
$T=0.5~\mathrm{s}$ and splitting it into $m=T/t_0$ independent
traces. By counting the number of electrons $N$ leaving the dot in
each trace and repeating the procedure for all $m$ sub-traces, the
distribution function $p_{t_0}(N)$ could be experimentally
determined. The experimental cumulants were then calculated directly
from the measured distribution function \cite{mathWorldCumulant:05}.
The time $t_0$ was chosen such that $\langle N \rangle \approx 3$.

\begin{figure}[htb]
\centering
 \includegraphics[width=\columnwidth]{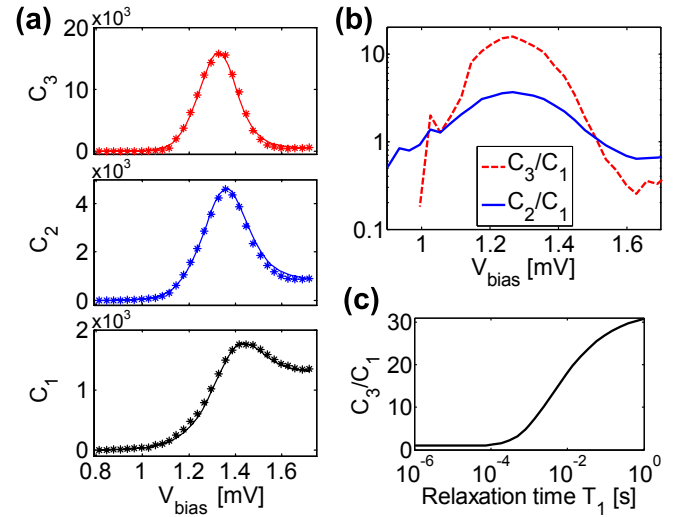}
 \caption{(Color online) (a) First, second and third cumulants of the distribution
 of charge fluctuations. The markers show values extracted from the experimental data, while
 the solid lines are calculated from the model given
 in the text. Fitting parameters are:
 $\Gamma_A = 1.6~\mathrm{kHz}$, $\Gamma_B = 20.5~\mathrm{kHz}$,
 $\Gamma_{\text{right}}^A = 4.6~\mathrm{kHz}$,
 $\Gamma_{\text{right}}^B = 310~\mathrm{Hz}$, $T_{1}=8~\mathrm{ms}$
 and $\mu_A-\mu_B =13~\mathrm{\mu eV}$. The electronic temperature was 400 mK.
 (b) Normalized cumulants $C_3/C_1$ and $C_2/C_1$
 versus bias voltage. The noise is clearly super-Poissonian in the
 central region of the graph. (c) Calculated maximal value of $C_3/C_1$ as a function
 of the relaxation time between the two states. The values are calculated
 by varying the relaxation time while keeping the other
 parameters to the values given by the fit shown in (a). The maximum value $C_3/C_1$
 extracted from the experimental data is $15.9$.
 }\label{fig:cumulants}
\end{figure}

Figure \ref{fig:cumulants}(a) shows the first three cumulants versus
voltage applied to the source lead. The points correspond to
experimental data, while the solid lines show the cumulants
calculated from the CGF of our model, with fitting parameters
$\Gamma_A = 1.6~\mathrm{kHz}$, $\Gamma_B = 20.5~\mathrm{kHz}$,
$\Gamma_{\text{right}}^A = 4.6~\mathrm{kHz}$,
$\Gamma_{\text{right}}^B = 310~\mathrm{Hz}$, $T_{1}=8~\mathrm{ms}$
and $\mu_A-\mu_B =13~\mathrm{\mu eV}$. The electronic temperature in
this measurement was 400 mK. The figure shows good agreement between
the model and the experimental data.

Figure \ref{fig:cumulants}(b) shows the normalized cumulants
$C_2/C_1$ and $C_3/C_1$ for the experimental data; we notice that
both the second and the third cumulants vastly exceed the first
cumulant when the Fermi level of the source lead is aligned with the
electrochemical potential of the dot ($V_{\mathrm{bias}} = 1.3~
\mathrm{mV}$). The noise is of super-Poissonian nature, as expected
from the bunching behavior of the electrons.

When the bias voltage is further increased ($V_{\mathrm{bias}} >
1.5~ \mathrm{mV}$), the source lead is no longer in resonance with
the electrochemical potential of the dot and the equilibrium
fluctuations between the source and the dot are suppressed. In this
regime, the measured charge fluctuations are due to a current
flowing through the dot. Electrons enter the dot from the source
lead and leave the dot through the drain lead. The blocking
mechanism is no longer effective and the transport process will
predominantly take place through state $S_A$, since the tunnel
coupling to the drain lead is stronger for this state
($\Gamma^A_{\text{right}} \gg \Gamma^B_{\text{right}}$). The
transport through the dot can essentially be described by a rate
equation, with one rate for electrons entering and another rate for
electrons leaving the dot. For such systems, it has been shown that
the Coulomb blockade will lead to an increase in correlation between
the tunneling electrons compared to a single-barrier structure,
giving sub-Poissonian noise \cite{davies:92, gustavsson:05}. The
effect is seen for $V_{\mathrm{bias}}>1.5~\mathrm{mV}$ in Fig
\ref{fig:cumulants}(b); both the second and third cumulants are
reduced compared to the first cumulant.

The value of $T_{1}=8~\mathrm{ms}$ obtained from fitting the
experimental data is of the same order of magnitude as previously
reported values for the spin relaxation time $T_1$. We stress that
the bunching of electrons and the super-Poissonian noise can only
exist if the relaxation time is at least as long as the inverse
tunneling time. This is demonstrated in Fig. \ref{fig:cumulants}(c),
which shows the maximum value obtained for the ratio $C_3/C_1$
calculated for different $T_{1}$ while keeping the rest of the
fitting parameters at the values given in the caption of Fig.
\ref{fig:cumulants}.

\section{Conclusion}
In this work, we have shown that a quantum point contact can be used
for measuring time-resolved transport through a weakly coupled
quantum dot. The detection method allows us to determine the
tunneling rates for electrons entering and leaving the dot
separately. Comparing the different tunneling rates, information
about the excited states and their relaxation times could be
extracted. We have shown that the framework of full counting
statistics together with time-resolved measurement techniques can be
used as a tool for extracting information about electron transport
properties of solid state systems.

\appendix
\section{Statistics of tunneling rates}\label{app:rates}
In the single-level regime, the process of an electron tunneling
into or out of the dot is described by the rate equation
\begin{equation}\label{app:eqRate}
    \dot{p}_{\mathrm{in/out}}(t) = - \Gamma_{\mathrm{in/out}}\,
    p_{\mathrm{in/out}}(t).
\end{equation}
Here, $p_{\mathrm{in/out}}(t)$ is the probability density for an
electron to tunnel into or out of the dot at a time $t$ after a
complementary event. Since the expressions for electrons entering
and leaving the dot are the same, we drop the subscripts ($in/out$)
and use the notations $p(t)$ and $\Gamma$ to describe either one of
the two processes. Solving the differential equation and normalizing
the resulting distribution gives
\begin{equation}\label{app:expDecay}
p(t) \mathrm{dt} = \Gamma \mathrm{e}^{-\Gamma t} \mathrm{dt}.
\end{equation}
Equation \ref{app:expDecay} is valid assuming non-degenerate states.
For the case of spin degeneracy, the rate for tunneling into the dot
should be multiplied with a factor of two if both spin states are
initially empty.

In the experiment, we measure a time trace containing a sequence of
tunneling times $\tau_k$, $k=1,2,3,\ldots$ To estimate $\Gamma$ and
its relative accuracy from such a sequence, we need to calculate the
probability distribution for extracting a certain value $\Gamma$,
given a fixed sequence of tunneling times. We start by dividing the
time axis into bins of width $\Delta \tau$ and number them with
$i=0,1,2,\ldots$ A tunneling event $\tau_{k}$ will be counted in bin
$i$ if $i\Delta \tau \leq \tau_{k} < (i+1)\Delta \tau$. Using Eq.
(\ref{app:expDecay}) and assuming $\Delta \tau \ll 1/\Gamma$, we
find that the probability to get a count in bin $i$ for a given
value of $\Gamma$ is equal to
\begin{equation}\label{app:eqPinOut}
 p(i|\Gamma) = \Gamma \Delta \tau
 \mathrm{e}^{-\Gamma \Delta \tau\,i}.
\end{equation}
A certain sequence $\{i_n\}$ is realized with probability
\begin{eqnarray}\label{app:eqSeq}
  p(\{i_n\}|\Gamma) &=& \prod_{n=1}^{N} \Gamma \Delta \tau
 \mathrm{e}^{-\Gamma \Delta \tau\,i_n} = (\Gamma \Delta \tau)^N \mathrm{e}^{-\Gamma \Delta \tau \sum_{n=1}^N
 i_n} \nonumber \\
 &=& (\Gamma \Delta \tau)^N \mathrm{e}^{-\Gamma \Delta \tau \sum_{i=0}^{\infty} n_i i} \nonumber \\
 &=& (\Gamma \Delta \tau)^N \mathrm{e}^{-\Gamma \Delta \tau N \langle i
 \rangle}.
\end{eqnarray}
Here, $n_i$ is the number of times an event falls into bin $i$,
$\sum_{i=0}^{\infty} n_i = N$ is the total number of events in the
trace and $\langle i \rangle = \frac{1}{N}\sum_{i=0}^{\infty} n_i i$
is the average of $i$. A certain set of bin occupations $\{n_i\}$
can be achieved with many different $\{i_n\}$-series, namely
$N!/\prod_{i=0}^{\infty} n_i !$. Assuming that they all occur with
the same probability $p(\{i_n\}|\Gamma)$, we find
\begin{equation}\label{app:eqSeq2}
  p(\{n_i\}|\Gamma) = \frac{N!}{\prod_{i=0}^{\infty} n_i !}
  (\Gamma \Delta \tau)^N \mathrm{e}^{-\Gamma \Delta \tau N \langle i
 \rangle}.
\end{equation}
This is our sampling distribution. For an estimate of $\Gamma$ we
use Bayes theorem
\begin{equation}\label{app:pToP}
  p(\Gamma|\{n_i\}) = p(\Gamma) \frac{p(\{n_i\}|\Gamma)}
  {p(\{n_i\})}.
\end{equation}
Because we have no information on the prior probabilities
$p(\Gamma)$ and $p(\{n_i\})$, the principle of indifference requires
them to be constants, giving
\begin{equation}\label{app:pGammaN1}
  p(\Gamma|\{n_i\}) = C \, (\Gamma \Delta \tau)^N \mathrm{e}^
  {-\Gamma \Delta \tau N \langle i \rangle},
\end{equation}
where $C$ is constant. Normalization $\int_0^{\infty}
p(\Gamma|\{n_i\}) \mathrm{d}\Gamma = 1$ leads to
\begin{eqnarray}\label{app:pGammaN2}
  p(\Gamma|\{n_i\}) &=& \frac{N^N \langle i \rangle^{N+1} \Delta \tau}{N!}
  (\Gamma \Delta \tau)^N \mathrm{e}^
  {-\Gamma \Delta \tau N \langle i \rangle} \nonumber \\
  &=& \frac{N^N}{N!} \langle \tau \rangle
  (\Gamma \langle \tau \rangle)^N \mathrm{e}^
  {-N \Gamma \langle \tau \rangle}.
\end{eqnarray}
The most likely value of $\Gamma$ is therefore $\Gamma^* = 1/\langle
\tau \rangle$. The relative accuracy of this estimate is given by
the width of the distribution. Setting $x=\Gamma \langle \tau
\rangle$ and evaluating the width at half maximum gives
\begin{eqnarray}\label{app:pGammaN3}
 x^N \mathrm{e}^{-x N} &=& \frac{1}{2} \mathrm{e}^{-N} \nonumber \\
 \Rightarrow \ln(x) &=& x -1 -\frac{1}{N} \ln(2).
\end{eqnarray}
For large $N$ we can expand $\ln(x)$ in a Taylor series around
$x=1$. Keeping only the first two terms, it follows
\begin{eqnarray}\label{app:pGammaN4}
 \frac{1}{2}(x-1)^2 &=& \frac{1}{N}\ln(2)  \nonumber \\
 \Rightarrow x &=& 1 \pm \sqrt{\frac{2 \ln(2)}{N}}.
\end{eqnarray}
Thus the relative accuracy is $\Delta \Gamma / \Gamma = \sqrt{2
\ln(2)/N}$.

\section{Fermi-distribution}\label{app:fermi}
With just one lead connected to the QD, we adopt the model
$\Gamma_{\mathrm{in}} = D_{n+1} \Gamma f(\Delta E/k_BT)$;
$\Gamma_{\mathrm{out}} = D_{n} \Gamma [1-f(\Delta E/k_BT)]$. Here,
$f$ is the Fermi distribution of the lead, $\Gamma$ is the tunnel
coupling between the QD and the lead, $D_n$ is the spin degeneracy
of the state with $n$ electrons on the QD and $\Delta E$ is the
energy difference between the electrochemical potential of the QD
and the Fermi level of the lead. The tunnel coupling is assumed to
be independent of energy within the small interval of interest.
Combining the equations gives
\begin{equation}\label{app2:fGamma}
 \frac{\Gamma_{\mathrm{in}}}{\Gamma_{\mathrm{out}}} =
 \frac{D_{n+1}}{D_n}\frac{f}{1-f}.
\end{equation}
In the following we will focus on the special case of non-degenerate
states, with $D_{n} = D_{n+1} = 1$.
To determine the Fermi distribution from a sequence of tunneling
events, we follow the lines of Appendix \ref{app:rates} and divide
the time axis into bins of width $\Delta \tau$. The bins are labeled
with $i=0,1,2,\ldots$, and an event will be counted in bin $i$ if
$i\Delta \tau \leq \tau_{\mathrm{in/out}} < (i+1)\Delta \tau$. We
collect the tunneling times in two sets of bins, one for electrons
tunneling into the dot ($\{n_i\}$), and one for electrons leaving
the dot ($\{m_i\}$). Assuming that the events corresponding to
tunneling in and tunneling out are uncorrelated and using the
results of Eq. (\ref{app:eqSeq2}), we get
\begin{eqnarray}\label{app2:2Gamma1}
 && p(\{n_i;m_i\}|\Gamma_{\mathrm{in}};\Gamma_{\mathrm{out}}) =
 p(\{n_i\}|\Gamma_{\mathrm{in}}) p(\{m_i\}|\Gamma_{\mathrm{out}}) \nonumber \\
 && \propto
 (\Gamma_{\mathrm{in}} \Gamma_{\mathrm{out}} \Delta\tau^2)^N
 \mathrm{e}^{-N(\Gamma_{\mathrm{in}} \langle \tau_{\mathrm{in}} \rangle +
 \Gamma_{\mathrm{out}} \langle \tau_{\mathrm{out}} \rangle)}.
\end{eqnarray}
The principle of indifference gives
\begin{eqnarray}\label{app2:2Gamma2}
 && p(\Gamma_{\mathrm{in}};\Gamma_{\mathrm{out}}|\{n_i;m_i\}) \propto \nonumber \\
 && ~~~ (\Gamma_{\mathrm{in}} \Gamma_{\mathrm{out}} \Delta\tau^2)^N
 \mathrm{e}^{-N(\Gamma_{\mathrm{in}} \langle \tau_{\mathrm{in}} \rangle +
 \Gamma_{\mathrm{out}} \langle \tau_{\mathrm{out}} \rangle)}.
\end{eqnarray}
Inserting the result of Eq. (\ref{app2:fGamma}) leads to
\begin{eqnarray}
 &&p(f;\Gamma_{\mathrm{out}}|\{n_i;m_i\}) \propto \nonumber \\
 &&\left(\frac{f}{1-f} \Gamma_{\mathrm{out}}^2 \Delta\tau^2 \right)^N
 \mathrm{e}^{-N \Gamma_{\mathrm{out}}\left(\frac{f}{1-f} \langle \tau_{\mathrm{in}} \rangle +
 \langle \tau_{\mathrm{out}} \rangle\right)}.
\end{eqnarray}
Marginalization gives
\begin{eqnarray}
 p(f|\{n_i;m_i\}) &&=
 \int_0^{\infty} p(f;\Gamma_{\mathrm{out}}|\{n_i;m_i\})
 \mathrm{d}\Gamma_{\mathrm{out}} \propto \nonumber \\
 && \frac{\left(\frac{f}{1-f}\right)^N (2N+1)!}
 {\left[\left(\frac{f}{1-f} \langle \tau_{\mathrm{in}} \rangle +
 \langle \tau_{\mathrm{out}} \rangle \right)N\right]^{2N+1}}.
\end{eqnarray}
The distribution has a maximum for $ \frac{f}{1-f} = \frac{ \langle
\tau_{\mathrm{out}} \rangle}{\langle \tau_{\mathrm{in}}
\rangle}\frac{N}{N+1}.$
In the limit of $N \rightarrow \infty$, the most likely $f$ is given
by
\begin{equation}
 f = \frac{\langle\tau_{\mathrm{out}} \rangle}
 {\langle \tau_{\mathrm{in}} \rangle + \langle \tau_{\mathrm{out}}\rangle}.
\end{equation}
We find the width of $p(f|\{n_i;m_i\})$ by approximating it around
its maximum by a Gaussian. The result is
\begin{equation}
 \sigma = f(1-f)\sqrt{2/N}.
\end{equation}


\bibliographystyle{apsrev}
\bibliography{SuperPoissonianNoise}

\end{document}